\documentclass[journal,onecolumn]{IEEEtran}

\ifCLASSINFOpdf
	\usepackage[pdftex]{graphicx}
    \graphicspath{{Figures/}}
    \DeclareGraphicsExtensions{.pdf}
\else
\fi
\usepackage{graphicx}
\usepackage{amsfonts}
\usepackage{amsmath}
\makeatletter
\def\Let@{\def\\{\notag\math@cr}}
\makeatother
\usepackage{colortbl}
\usepackage{cite}
\usepackage{booktabs}
\usepackage{multirow}
\usepackage{algorithm} 
\usepackage{algorithmic}
\usepackage[shortlabels]{enumitem}
\usepackage{accents}
\usepackage{color}
\usepackage{amsmath}
\usepackage{amssymb}
\usepackage{url}
\usepackage{hyphenat}
\usepackage{comment}
\usepackage{upgreek}
\usepackage{subfigure}
\usepackage{chngcntr}
\begin{document}

\title{\color{black}Tracking Power System Events with Accuracy-Based PMU Adaptive Reporting Rate}

\author{Guglielmo Frigo, 
Paolo Attilio Pegoraro,
Sergio Toscani\\
\vspace{0.2cm}
\thanks{
G. Frigo is with the Electrical Energy and Power Lab, Federal Institute of Metrology, Bern-Wabern, Switzerland (email: guglielmo.frigo@metas.ch).

P. A. Pegoraro is with the Department of Electrical and Electronic Engineering of the University of Cagliari, Piazza d'Armi, 09123 Cagliari, Italy (email: paolo.pegoraro@unica.it).

S. Toscani is with Dipartimento di Elettronica, Informazione e Bioingegneria of the Politecnico di Milano, Milano, Italy (email: sergio.toscani@polimi.it).} }

\maketitle
{\color{black}
\begin{abstract}
Fast dynamics and transient events are becoming more and more frequent in power systems, due to the high penetration of renewable energy sources and the consequent lack of inertia. In this scenario, Phasor Measurement Units (PMUs) are expected to track the monitored quantities. Such functionality is related not only to the PMU accuracy (as per the IEC/IEEE 60255-118-1 standard) but also to the PMU reporting rate (RR). High RRs allow tracking fast dynamics, but produce many redundant measurement data in normal conditions. In view of an effective tradeoff, the present paper proposes an adaptive RR mechanism based on a real-time selection of the measurements, with the target of preserving the information content while reducing the data rate. The proposed method has been tested considering real-world datasets and applied to four different PMU algorithms. The results prove the method effectiveness in reducing the average data throughput as well as its scalability at PMU concentrator or storage level.

\end{abstract}

}


\section{Introduction}
Power systems are experiencing an ever-increasing penetration of distributed generation from renewable energy sources (RES) \cite{Arnold-ProcIEEE-2011}. Due to their inherent volatility and lack of rotational inertia, RES are more prone to sudden variations, which may result in regional or even systemic contingencies \cite{Paolone-etAl-PSCC-2020,Turkey-2015,Australia-2016}. 

{\color{black}In this scenario, the measurement infrastructure shall guarantee a prompt and capillary control, e.g. by means of Phasor Measurement Units (PMUs) \cite{Rietveld-etAl-TIM-2015,DeLaRee-etAl-TSG-2010,Moraes-etAl-TSG-2012}. Indeed, PMUs produce time-stamped measurements of synchrophasor, frequency, and rate of change of frequency (ROCOF), enabling to compare the power system state in remote locations.} 
Moreover, PMUs are characterised by a reduced latency of few tens of ms, and a reporting rate (RR) of tens frames per second (fps).

It is worth noticing that PMUs were originally conceived for large transmission networks, in order to measure phasor and frequency deviations between remote nodes, assuming quasi-steady state conditions. The objective was to provide a snapshot of the system state at a given reporting time instant\cite{Zhong-etAl-TPWRS-2005,Kamwa-etAl-TPWRS-2013}, rather than describing a time evolution. 

Nowadays, the challenging conditions induced by the reduced inertia have pushed for the development of more and more accurate estimation algorithms, capable of dealing with faster dynamics. However, this is not sufficient to properly identify trends in consecutive measurements \cite{Chakir-etAl-TPWRD-2014}. Indeed, this capability does not just depend on the estimation accuracy of a given algorithm, on the analog front end \cite{Ma-etAl-TSG-2010,Follum-etAl-TPWRS-2017} or the employed transducer \cite{FerPegTosCVT}, but it is strongly affected by the configuration of the PMU, particularly by its RR, and by how measurement data are employed. 

{\color{black}
The relationship between dynamics monitoring and PMU RR was investigated in the literature. In \cite{2017TIM_Pegoraro_IoT}, within an internet of things framework, the RR of a PMU is adaptively tuned based on root mean square (rms) variation of voltage measurements to follow dynamics and feed a variable rate state estimator. In \cite{2022SGSMA_Frigo_TrackingEvents}, the impact of PMU accuracy and RR with different interpolation techniques is investigated.
The ideal tracking should be 
able to measure, interpolate or predict the quantity of interest in each relevant time instant. For this reason, \cite{2022SGSMA_Frigo_TrackingEvents} introduces the concept of tracking error index, intended to summarise the usefulness of available measurements for describing a time evolution.

In this paper, 
we introduce a strategy to reduce the RR, and thus the PMU data throughput, depending on the comparison between measured and predicted value in each potential reporting instant. Following this approach, it is possible to prevent a measurement from being transmitted when the tracking quality at the receiver is not significantly reduced. 
Until the tracking quality is sufficient, there is no need to transmit new redundant data. In other words, the measurement is transmitted only if it carries a significant amount of independent information with respect to what is already available. 
Differently from \cite{2017TIM_Pegoraro_IoT}, the proposed method focuses on the application viewpoint: the objective is to guarantee a sufficient tracking quality while reducing the data throughput. 
It can be applied, with possibly different configurations, at PMU, phasor data concentrator (PDC), application, and even data storage level, aiming at keeping relevant information in a concise way when and where needed. The proposed method is simple, real-time and keeps into account accuracy targets according to the specific requirements. {\color{black}Furthermore, it is complementary to compression-based strategies, like in \cite{2016TSG_Gadde_PMUDataCompression,2020TII_Wang_Synchrophasor_Compression}, which are typically post processing stages for storage purposes.}

{\color{black}The paper is organised as follows. Section \ref{sec:algorithm} introduces the concept of tracking index and its application to reduce the RR. Section \ref{sec:case_studies} describes the case studies derived from real-world events and used as testbench for the proposed method. Section~\ref{sec:assume} briefly presents the adopted PMU algorithms. Section~\ref{sec:results} illustrates the performed simulations and discusses the obtained results. Finally, Section \ref{sec:concl} gives some closing remarks and possible future lines of research.}


\section{Proposed Method}
\label{sec:algorithm}

As shown in \cite{2022SGSMA_Frigo_TrackingEvents}, the tracking capability of a PMU does not descend directly from its point-wise accuracy, as this is defined only in correspondence of the reporting time instants. By interpolating or extrapolating new measurement values, it is possible to introduce specific indices for tracking error. 

\subsection{Tracking Indices}
\label{sec:tracking_index}
The most relevant metrics to evaluate PMU errors are:}
\paragraph{Total Vector Error (TVE)} {\color{black}For a given time instant $t$, it represents the relative distance (in percentage) in the complex plane} between estimated and reference synchrophasor:
    {\color{black}
    \begin{equation}
        \textrm{TVE}(t) = 100 \cdot\frac{|\widehat{\Bar{X}}(t)-\Bar{X}(t)|}{|\Bar{X}(t)|}
        \label{eq:TVE}
    \end{equation}
    where $t$ is the measurement time instant, $\Bar{X}(t)$ is the synchrophasor at time $t$, and the symbol $\;\widehat{}\;$ indicates a measured value hereinafter.
    }
    {\color{black}TVE can be straightforwardly generalised to evaluate the relative deviation between two generic phasors and it will be used also with this meaning in the following.}
    \paragraph{Frequency error (FE)} It is the difference between measured and reference frequency at time $t$, hence
    \begin{equation}
        \textrm{FE}(t) =  \widehat{f}_1(t)-f_1(t)
    \end{equation}
    {\color{black}with subscript `$1$' underlining that the frequency of the fundamental component is considered.} In the following, FE will be expressed in mHz.
    \paragraph{ROCOF error (RFE)} It corresponds to the difference between measured and reference ROCOF at time $t$, namely
    \begin{equation}
        \textrm{RFE}(t) =  \widehat{\textrm{ROCOF}}(t)-\textrm{ROCOF}(t)
    \end{equation}
    that will be expressed in Hz/s.
    
    {\color{black}
    In \cite{2022SGSMA_Frigo_TrackingEvents}, the so-called Tracking Error  index ($\textrm{TrE}$) is introduced. In this paper, a more general definition of $\textrm{TrE}$ is adopted, as described below.
    Three $\textrm{TrE}$s are considered, namely $\textrm{TrE}_{\textrm{TVE}}$, $\textrm{TrE}_{\textrm{FE}}$, $\textrm{TrE}_{\textrm{RFE}}$, respectively for quantifying the tracking accuracy of phasor, frequency and ROCOF. Only the first one will be explicitly defined, since the other two can be derived straightforwardly. Measurement units are those of the corresponding error metrics.
    Concerning the phasor
    \begin{equation}
        \textrm{TrE}_\textrm{TVE} = \sqrt{\frac{1}{N}\sum_{n=0}^{N-1}\frac{|\widetilde{\Bar{X}}(nT_s)-\Bar{X}(nT_s)|}{|\Bar{X}(nT_s)|}}.
        \label{eq:TrE_TVE}
    \end{equation}
    
    $T_s$ is the sampling interval, corresponding to the maximum available data rate (the sample rate $f_s$), namely the closest approximation of the continuous time domain. $N$ is the number of samples in the considered time window (the timescale is shifted so that the first sample is located in $t=0\,\textrm{s}$). The symbol  $\;\widetilde{}\;$  indicates the measured or reconstructed value, i.e.
    \begin{equation}
        \widetilde{\Bar{X}}(nT_s)=
        \begin{cases}
            \widehat{\Bar{X}}(nT_s)\quad\;\,\textrm{for $nT_s=t_i$} \\
            \Bar{X}^p(nT_s)\quad\textrm{for $nT_s\neq t_i$}
        \end{cases}
    \end{equation}
    $t_i=k_iT_s$, with $k_i\in\mathcal{K} \subset\left\{0,\ldots,N-1\right\}$ being the sample index corresponding to the $i$th available measurement. $\Bar{X}^p(nT_s)$ is the reconstructed synchrophasor at the instant $nT_s$, which does not correspond to a measurement instant. For the sake of simplicity and without loss of generality, in the following each PMU reporting instant corresponds to a sampling instant.

    The TrE is thus defined as the rms value of the deviation between the reference quantity, sampled with rate $f_s$, and its reconstruction, obtained from a set of measurements and provided with rate $f_s$ through a proper prediction or interpolation technique. In this paper, prediction only is considered (thus exploiting just past measurements) because we focus on real-time measurement manipulation, while interpolation unavoidably increases latency. Prediction can be performed with many different algorithms. In the following, phasor extrapolation from the last measurement instant is considered, therefore
    \begin{eqnarray}
    \label{eq:phasor_prediction}
            &&\widetilde{\bar{X}}(t_{i(n)}+(n-k_{i(n)})T_s) =\widehat{\Bar{X}}(t_{i(n)})\cdot\\ 
            &&\cdot
            e^{j\left\{2\pi(\widehat{f}_1(t_{i(n)})-f_0)(n-k_{i(n)})T_s+\pi     \widehat{\textrm{ROCOF}}(t_{i(n)})(n-k_{i(n)})^2T_s^2\right\}} \nonumber 
    \end{eqnarray}
    where $t_{i(n)}=k_{i(n)}T_s$ is the last available or considered measurement instant before time $nT_s$ (i.e., $i(n)=\arg \max_i\left\{k_i\;|\;k_i\leq n\right\}$
    ). The most recent synchrophasor, frequency and ROCOF measurements, among those available at a given processing stage, are thus used to compute the predicted synchrophasor. This implies that} 
    phasor amplitude is simply repeated every $T_s$ until a new measurement is available, while phasor phase-angle is extended to future instants using the latest frequency and ROCOF estimates. 
    {\color{black} $\textrm{TrE}_\textrm{FE}$ and $\textrm{TrE}_\textrm{RFE}$ are defined analogously to \eqref{eq:TrE_TVE} as the rms FE and RFE, respectively, between predicted and reference quantities in the considered interval, using time step $T_s$.}
    Following the same approach, frequency extrapolation is performed as
    \begin{align}
        \widetilde{f}_1(t_{i(n)}+(n-k_{i(n)})T_s)&=\widehat{f}_1(t_{i(n)})+\\
        &+\widehat{\textrm{ROCOF}}(t_{i(n)})(n-k_{i(n)})T_s
        \label{eq:freq_prediction}
    \end{align}
    while ROCOF is obtained simply by holding the most recent value until the next one is available, i.e.
    \begin{equation}
      \widetilde{\textrm{ROCOF}}(t_{i(n)}+(n-k_{i(n)})T_s)=\widehat{\textrm{ROCOF}}(t_{i(n)}).
      \label{eq:rocof_prediction}
    \end{equation}

{\color{black}
\subsection{Adaptive Measurement Decimation}   

Tracking error indices $\textrm{TrE}_\textrm{TVE}$, $\textrm{TrE}_\textrm{FE}$, and $\textrm{TrE}_\textrm{RFE}$ 
are intended to assess the capability to follow the time evolution of synchrophasor, frequency and ROCOF, respectively, during an event, by exploiting the available measurements $\widehat{\Bar{X}}(t_{i})$, $\widehat{f}_1(t_{i})$, $\widehat{\textrm{ROCOF}}(t_{i})$, with $i\;|\;k_i\in\left\{0,k_{i(1)},\ldots,k_{i(N-1)}\right\}$.
For this reason, they provide an indication of the available information (possibility to catch electrical signal dynamics) and thus they depend on both the point-wise PMU accuracy and the location of the time instants $k_iT_s$, other than the method adopted for prediction. 

Reporting time instants $t_i$ are intrinsically sparse with respect to the sampling instants, because RR in a PMU cannot be higher than a few hundred frames per second. 
The capability to predict synchrophasor, frequency and ROCOF at unmonitored time instants is thus limited by measurement accuracy and rate when coupled to each phenomenon timescale. It is thus impossible to define a unique RR that is optimal for all the conditions of interest. For example, in a steady-state scenario, few measurements per second are enough to get all the needed information, but when dynamics occur, the RR should increase accordingly to track the variations of the monitored quantities. On the other hand, a constantly high RR represents a waste of resources (bandwidth, processing, storage, etc.) since it produces a transfer of heavily redundant information, and might even jeopardise applications responsiveness.

In the proposed method, the PMU algorithm runs at high, fixed rate $\textrm{RR}_{\textrm{in}}$ (corresponding to the interval $T_{\textrm{in}}$) but a measurement is transmitted only if 
useful under the circumstances at hand. An adaptive RR 
is able to better respond to dynamics while reducing data throughput. 

This method is applicable every time a PMU measurement stream can or needs to be decimated, without adding significant latency. The algorithm works in real-time on a sequence of PMU measurements. Assuming that $T_{\textrm{in}}=rT_s$, we consider a selection strategy for each incoming triplet ($^T$~is the transpose operator)\footnote{The index scanning the reporting instants is $h=0,1,\ldots,\left\lfloor\frac{N}{r}\right\rfloor$}
\begin{equation}
 \mathbf{m}(hrT_s)=\left[\widehat{\Bar{X}}(hrT_s),\widehat{f}_1(hrT_s),\widehat{\textrm{ROCOF}}(hrT_s)\right]^T   
\end{equation}
that issues the following decimated output sequence:
\begin{align}
\label{eq:decimation_rule}
 &\mathbf{m}^{\textrm{out}}[i]=\mathbf{m}(t_{i})\\
 &=
        \begin{cases}
            \mathbf{m}(h(0)rT_s)=\mathbf{m}(0)\quad\textrm{for $i=0$}\\
            \\
            \!\!\!\begin{array}{ll}
                  \mathbf{m}(h(i)rT_s)&\!\textrm{where}\; h(i)=  
                  \arg \min_h \left\{h \,|\, h > h(i-1)\,\&\right.\\ &\left.\|\boldsymbol{\upepsilon}_i(\mathbf{m}(hrT_s))\|_\infty > 1  \right\}
            \end{array}
        \end{cases}
\end{align}
where $h(i)$ is the reporting index $h$ associated with the $i$th selected measurement set, 
$\|\cdot\|_\infty$ is the infinity norm, while  $\boldsymbol{\upepsilon}_i(\cdot)$ is a vector function defined as:
\begin{align}
    \boldsymbol{\upepsilon}_i(\mathbf{m}(hrT_s))=
    \arraycolsep=1.2pt\def\arraystretch{1.7}
    \begin{bmatrix}
      \frac{|\Bar{X}^p(hrT_s)-\widehat{\Bar{X}}(h(i-1)rT_s)|}{\Delta_{\textrm{TVE}}|\widehat{\Bar{X}}(h(i-1)rT_s)|}\\
      \frac{f_1^p(hrT_s)-\widehat{f}_1(h(i-1)rT_s)}{\Delta_{\textrm{FE}}}\\
      \frac{\textrm{ROCOF}^p(hrT_s)-\widehat{\textrm{ROCOF}}(h(i-1)rT_s)}{\Delta_{\textrm{RFE}}}
    \end{bmatrix}
\end{align}
$\Bar{X}^p(hrT_s)$, $f_1^p(hrT_s)$, and $\textrm{ROCOF}^p(hrT_s)$ are the predicted synchrophasor, frequency and ROCOF values, respectively, at $t=hrT_s$, which are obtained as mentioned above, i.e., applying \eqref{eq:phasor_prediction}, \eqref{eq:freq_prediction}, and \eqref{eq:rocof_prediction} considering the instant $t_{i-1}=h(i-1)rT_s$, corresponding to the last selected measurement set $\mathbf{m}^{\textrm{out}}[i-1]$. Parameters $\Delta_{\textrm{TVE}}$, $\Delta_{\textrm{FE}}$ and $\Delta_{\textrm{RFE}}$ are normalisation factors that correspond to the thresholds imposed to phasor, frequency and ROCOF prediction deviations, respectively.

From an intuitive point of view, the algorithm keeps and transmits only the measurements considered as relevant with the following approach: at every instant $hrT_s$, it computes the predicted values starting from the last selected triplet of measured values (index $i-1$) and compares it with the current triplet $\mathbf{m}(hrT_s)$ in terms of TVE, FE, and RFE. If at least a deviation exceeds the corresponding threshold, $\mathbf{m}(hrT_s)$ is kept and becomes $\mathbf{m}^{\textrm{out}}[i]$. Otherwise, data is discarded and the process moves on iteratively, further extending the time horizon by $T_{\textrm{in}}$ and updating the prediction.

The same strategy can be applied also to individual quantities, e.g., to frequency measurement only, thus extracting from the measurement flow at RR a subset of relevant measurements. This can help in further reducing the amount of data, but, in the following, only the global rule defined by \eqref{eq:decimation_rule} (i.e., the thresholds on the three parameters are checked as if they were conditions in `OR') is discussed and tested.
}

\section{Case Studies}
\label{sec:case_studies}
\begin{figure}
    \centering
    \includegraphics[width=1\columnwidth]{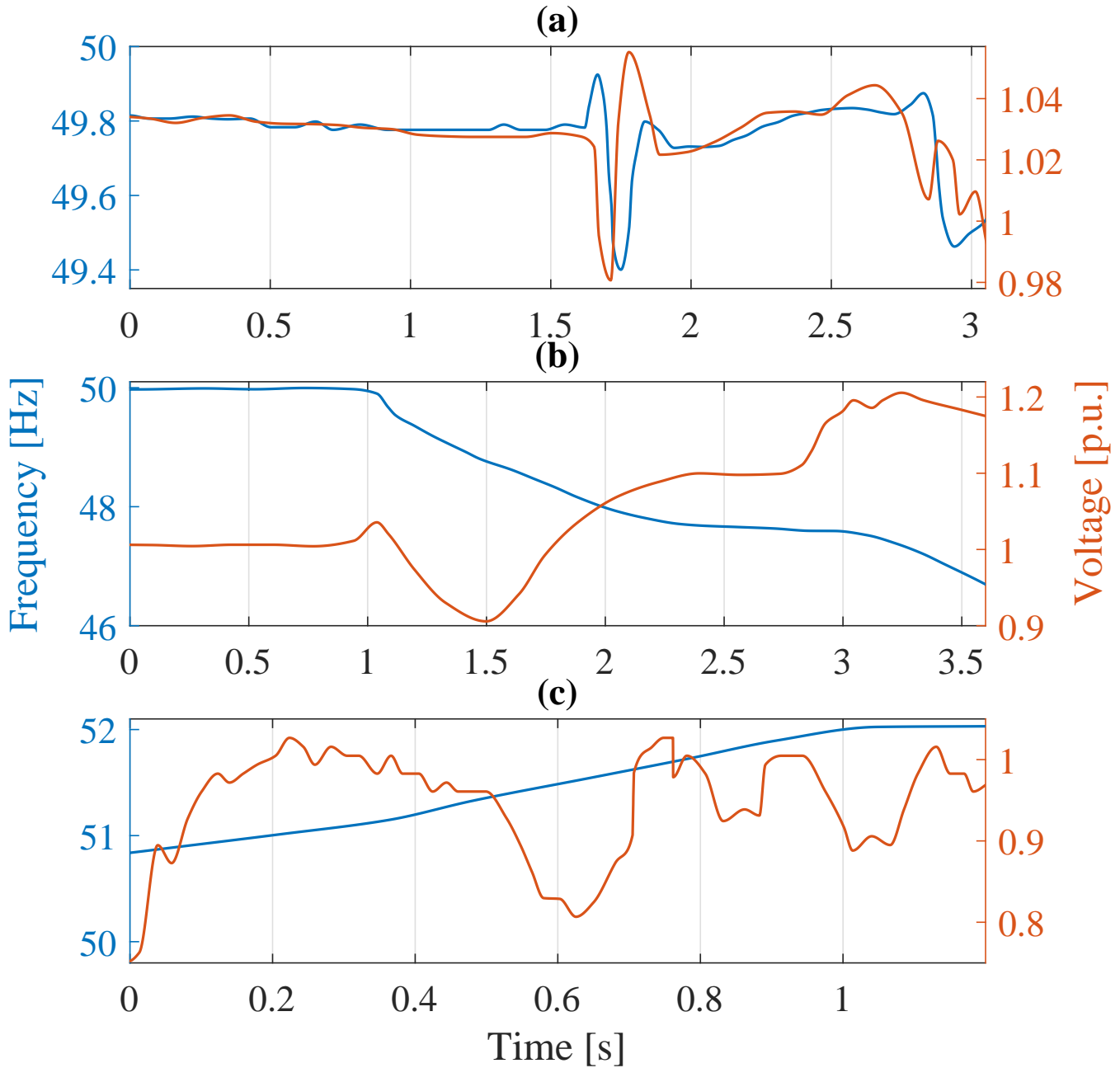}
    \caption{{\color{black}Instantaneous frequency (blue) and voltage amplitude (red) as function of time in the three considered short-duration case studies: Australia 2016 in (a), Arizona 2011 in (b), and Turkey 2015 in (c).}}
    \label{fig:case_studies_short}
    \vspace{-.25cm}
\end{figure}

\begin{figure}
    \centering
    \includegraphics[width=\columnwidth]{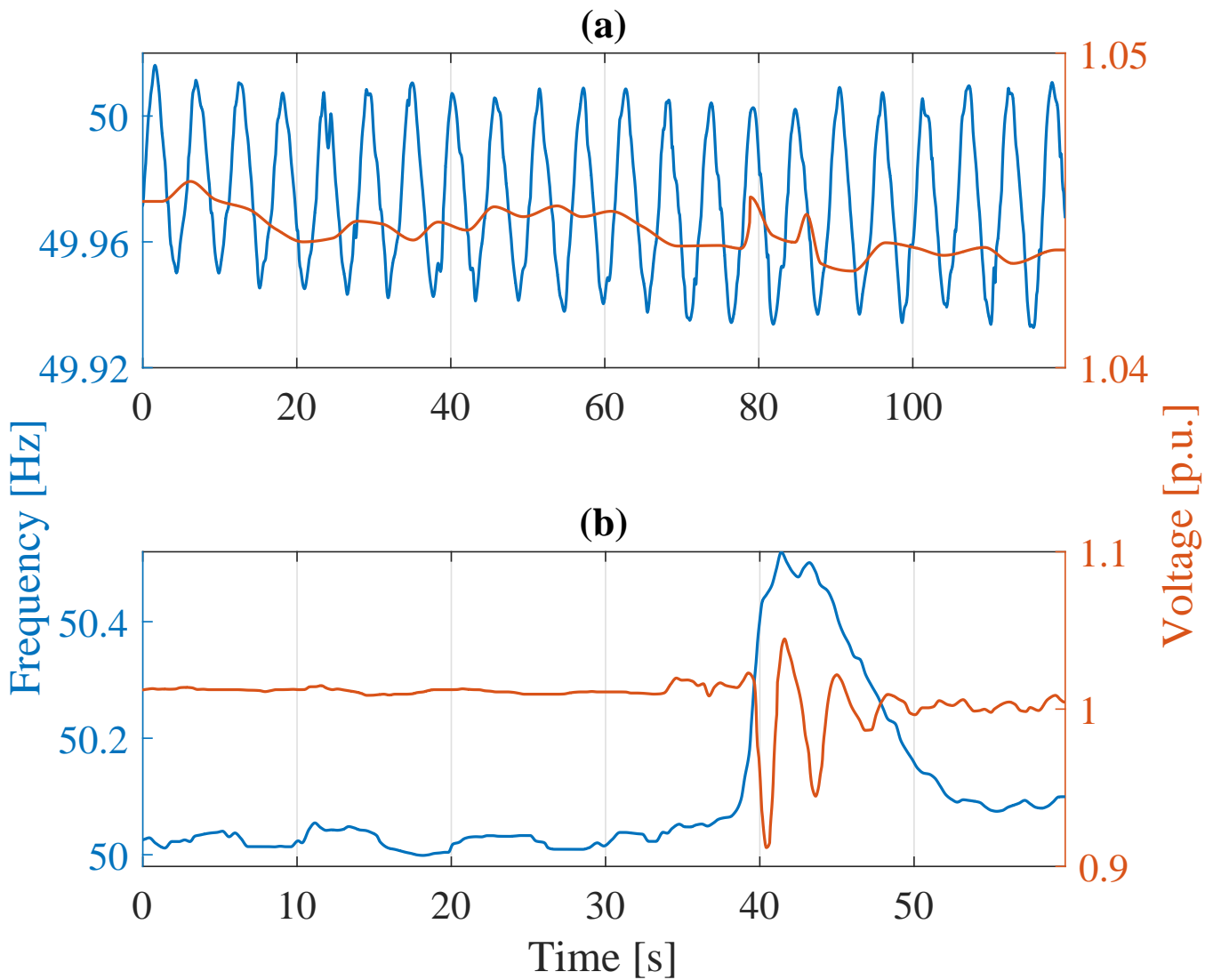}
    \caption{{\color{black}Instantaneous frequency (blue) and voltage amplitude (red) as function of time in the two considered long-duration case studies: Florida 2019 in (a) and Croatia 2021 in (b).}}
    \label{fig:case_studies_long}
\end{figure}

{\color{black}
In this Section, we briefly introduce the case studies used for evaluating the performance that can be reached by adopting the proposed adaptive RR algorithm.}
Such case studies have been derived from official reports of contingencies and transient events, as published by transmission network operators and regulating agencies. This enables assessing the performance under real-world scenarios.
{\color{black} In particular, two different sets of case studies have been considered. The first one, reported in Fig. \ref{fig:case_studies_short}, comprises three short segments of contingencies intended to demonstrate the proposed method capability of promptly adapting the RR based on the current conditions. The second set, instead, is presented in Fig. \ref{fig:case_studies_long} and consists of longer duration events, which allow for evaluating the proposed method advantages in terms of RR optimisation also in the presence of normal or slowly-varying conditions.

The following paragraphs provide a short description of each case study. More precisely, \textit{a)}, \textit{b)}, and \textit{c)} refer to the first set, whereas \textit{d)} and \textit{e)} refer to the second one. In this context, we focus on limited portions of signals that are particularly relevant to the proposed method, but further information can be found in the corresponding official reports.
}

{\color{black}\paragraph{Australia 2016} The first case study refers to the South Australia blackout, occurred on September 28, 2016 \cite{Australia-2016}. Extreme weather conditions caused the trip of three transmission lines and initiated a sequence of 6 faults. Simultaneously, 9 wind farms were forced to a sustained generation reduction as a protection feature activated. The combination of such adverse factors produced a sudden separation of the South Australian system, thus resulting in frequency instability. With an average ROCOF of $-6.25\,\textrm{Hz/s}$, the remaining generation facilities were tripped in less than $1\,\textrm{s}$ and caused the blackout for the entire region.}

{\color{black}Fig. \ref{fig:case_studies_short}(a) shows the time profiles of voltage magnitude and frequency as measured at the node of Robertstown in red and blue, respectively. The rated voltage and frequency were equal to 275 kV and 50 Hz, respectively. After nearly 1.5 s, both the profiles exhibit a sudden transition from normal to transient conditions, characterised by fast dynamics and oscillations. 
}

\paragraph{Arizona 2011} {\color{black}The second case study} refers to a sequence of outages that hit the Pacific Southwest power system on September 8, 2011 \cite{Arizona-2011}. The event had an overall duration of 11 minutes. It was initiated by a series of unfortunate factors, as the opening of a transmission line in correspondence of a demand peak, and it ended with the system collapse due to the irreparable imbalance between generation and demand. In particular, the last part has been considered, when a rapid drop of both voltage and frequency produces the unavoidable trip of both loads and generators.

{\color{black}Fig. \ref{fig:case_studies_short}(b) reports the instantaneous voltage amplitude and frequency in red and blue, respectively. After nearly $1\,\textrm{s}$, a first separation of the system causes a drop in both profiles. Despite a partial mitigation provided by under-frequency load shedding, after $3\,\textrm s$ the synchronous generators are tripped and the frequency collapse starts again.} 
The original dataset refers to a transmission network with rated voltage and frequency equal to $300\,\textrm{kV}$ and $60\,\textrm{Hz}$, respectively. Without loss of generality, in the following, frequency has been scaled to {50\,\textrm{Hz}}. First, this guarantees the comparability of the results with the other cases, since the same PMU algorithms can be employed. Second, the obtained frequency profile 
still represents a plausible scenario for a power system outage.

\paragraph{Turkey 2015} The third case study refers to the blackout occurred in Turkey on March 31, 2015. As reported in \cite{Turkey-2015}, the trip of an overloaded line produced the separation into an Eastern and Western subsystem. In the latter one, the sudden power deficit of nearly $21\,\%$ caused the loss of synchronism with the Central Europe system: after $1\,\textrm{s}$, the three interconnection lines with the Bulgarian and Greek grids were also tripped. As the frequency was rapidly dropping from 49 to $48.4\,\textrm{Hz}$, the under-frequency load shedding relays disconnected about $4800\,\textrm{MW}$ of load, thus reducing the ROCOF. Once reached the frequency limit of $47.5\,\textrm{Hz}$, though, the loss of several synchronous generators caused the collapse of the entire Western subsystem. 

{\color{black}The contingency has an overall duration of $10\,\textrm{s}$, but the present analysis focuses on a shorter portion, shown in Fig.~\ref{fig:case_studies_short}(c) and characterised by inconsistent and uncorrelated dynamics for frequency and voltage. More precisely, the frequency experiences a steady, yet progressively increasing, linear ramp, whereas the voltage presents significant modulation. For the sake of completeness, it is worth mentioning that this specific event is triggered by the opening of the Ataturk-Yesilhisar Kuzey $400\,\textrm{kV}$ transmission line.}


{\color{black}\paragraph{Florida 2019} The fourth case study refers to a forced oscillation event as recorded in the US Eastern Interconnection on January 11, 2019 \cite{Florida-2019}. Due to a failure in the control of a steam turbine generation unit, an oscillation was observed in the Florida network for an overall duration of approximately 18 minutes. The event was characterised by a dominant frequency around 0.25 Hz with a near-zero dampening ratio. In this case, the rated voltage and frequency are equal to 500 kV and 60 Hz, respectively. Also in this case, we transposed the frequency of the synthetic waveforms at 50 Hz.

Fig. \ref{fig:case_studies_long}(a) presents the voltage and frequency time profiles as acquired by Frequency Disturbance Recorders \cite{Liu-etAl2021-Access} in red and blue, respectively. For the sake of clarity, the analysis is limited to a segment of two minutes but similar considerations hold for the entire event duration. It is interesting to observe how the frequency profile is affected by a sinusoidal modulation whose frequency and depth can be approximated to 0.25 Hz and 50 mHz, respectively. The voltage profile, instead, keeps quite stable around the pre-event value.}

{\color{black}\paragraph{Croatia 2021} The fifth case study refers to the system separation occurred in Croatia on 8 January 2021 \cite{Croatia-2021}. Due to a series of cascaded trips of transmission network elements (initiated in the substation Ernestinovo, HR), the Continental Europe Synchronous Area was split in two areas, which in first approximation could be referred to as North-West and South-East. The immediate reaction by means of automated and manual countermeasures allowed to restore the normal operation in nearly one hour, without major consequences in terms of damages or shed loads.

Fig. \ref{fig:case_studies_long}(b) presents the voltage and frequency time evolution as recorded at the substation of Hamitabat in Turkey in red and blue, respectively. In this case, the rated voltage is 400~kV and the nominal system frequency is equal to 50 Hz. For the sake of brevity, the analysis focuses on a limited segment of just one minute. The system separation initiates after around 40 s. Then, both the profiles exhibit different transient behaviours lasting nearly 10 s. The frequency rapidly ramps up to 50.5 Hz, before initiating a slow decaying trend. The voltage instead is affected by periodically repeated dips, whose depth is progressively decaying.}

{\color{black}\paragraph*{Test cases' significance} The dynamics included in the selected case studies do not coincide in terms of traditional PMU metrics (e.g. maximum ROCOF or response time), neither represent a combination of test conditions included in the IEC/IEEE 60255-118-1 standard (IEC Std) compliance verification \cite{IEC-IEEE-60255-118-1-2018}. Their main common feature is the presence of either fast transients or oscillatory trends. From a power system control perspective, both these aspects need to be properly detected and monitored and thus represent the ideal testbench for the assessment of the dynamic tracking capability achievable with PMUs.}

For each test case, data points of voltage amplitude and frequency have been time interpolated through shape-preserving piecewise cubic Hermite polynomials. The resulting interpolations feature continuous derivatives and monotonic trends between couples of points, so that the shape of the original data is preserved.
Phase angle and ROCOF are represented with piecewise polynomials that can be respectively obtained by time integrating and differentiating the analytical interpolation of frequency data. The expressions are employed as ground-truth values for the positive sequence synchrophasor, frequency and ROCOF to be estimated. Of course, this does not mean that they exactly correspond to the original event: they are affected by several uncertainty contributions due, for example, to the transducers, the data acquisition systems, the estimation algorithms as well as the adopted interpolation approach. In any case, the obtained trends can be still considered as representative of the dynamic scenarios that may occur in real power systems.

Assuming purely positive sequence voltages, the samples of the three-phase waveforms, which are the input of the synchrophasor estimation algorithms, can be easily computed from the interpolated amplitude and phase data; $10\,\textrm{kHz}$ sampling rate $f_s$ was adopted, thus $M=200$ samples per cycle at the considered $f_0=50\,\textrm{Hz}$ rated frequency. Previous hypotheses enable focusing only on the tracking capability of the considered PMU algorithms, without the potentially detrimental effect of other disturbances.

\section{Test Assumptions}
\label{sec:assume}
{\color{black}
The implementation of the proposed adaptive decimation approach firstly requires assuming a RR value for the incoming data. 
For the sake of compatibility, $\textrm{RR}_{\textrm{in}}=100\,\textrm{fps}$, thus corresponding
to the highest standardised RR for $50\,\textrm{Hz}$ systems \cite{IEC-IEEE-60255-118-1-2018}, is used.
The values adopted for the thresholds $\Delta_{\textrm{TVE}}$, $\Delta_{\textrm{FE}}$ and $\Delta_{\textrm{RFE}}$ should be tuned to reach the desired tradeoff between tracking performance and amount of data. An appropriate choice should consider also the accuracy of the raw PMU measurement data, which depends on the adopted PMU estimation algorithm, but also on the class of the instrument transformer. More specifically, it makes no sense to further reduce the threshold values when they are still negligible with respect to the uncertainty of the corresponding input estimates. According to this consideration, in the tests $\Delta_{\textrm{TVE}}$ (the threshold value for the TVE) was set to $10^{-3}$ ($0.1\,\%$), which is one tenth of the accuracy requirement for PMU algorithms under steady state conditions \cite{IEC-IEEE-60255-118-1-2018}, reminding that they mostly behave much better. It is worth stressing that the selected value is also significantly lower than that derived from the accuracy class of the typical instrument transformers for power systems applications. On the other hand, instrument transformer errors are mainly systematic as far as synchrophasor measurement is concerned and thus they do not affect the proposed measurement selection procedure in a significant way. Furthermore, under slow dynamics, the instrument transformer has negligible impact on the accuracy of frequency and ROCOF estimates, which is bounded by the performance of the PMU algorithm. Therefore, $1\,\textrm{mHz}$ was selected as FE threshold $\Delta_{\textrm{FE}}$, namely one fifth of the error requirement for PMU algorithms undergoing steady-state tests without disturbances. Finally, $\Delta_{\textrm{RFE}}=0.07\,\textrm{Hz/s}$ was chosen, thus almost six times lower than the maximum RFE value for P-class compliant algorithms during steady-state conditions.

The method has been applied to decimate the outputs of four different PMU algorithms fed with the waveforms described in Section \ref{sec:case_studies} in order to evaluate their capability to reduce data transfer with limited impact on tracking performance.
All the algorithms are P-class compliant}, thus designed for achieving fast response. However, they are based on very different estimation approaches, which are briefly described in the following:

    \paragraph{P-class reference algorithm (P-IEC)} P-class algorithm proposed by the IEC Std \cite{IEC-IEEE-60255-118-1-2018}. The synchrophasor extraction method consists of demodulation through a mixer stage tuned at the rated frequency $f_0$ and filtering by a $N_w = 2M-1$ sample triangular window (with $M$ the number of samples in a nominal signal cycle). 
    The synchrophasor is computed for each system phase at an internal frequency, which can be higher than the considered $\textrm{RR}$ (here $f_s$ is used, that is a sample-by-sample estimate); positive-sequence synchrophasor is computed applying the Fortescue transformation. Frequency and ROCOF measurements are obtained through symmetric first and second order discrete-time derivatives of the phase-angle of the positive-sequence synchrophasor, respectively. Measured frequency is also exploited in order to compensate for the attenuation introduced by the filter under off-nominal frequency conditions. The different RRs are obtained subsampling the internal measurements.
    \paragraph{Iterative-Interpolated DFT (i-IpDFT)} The algorithm \cite{2017TIM_Derviskadic_P+M} is based on an enhanced version of the three-point IpDFT, thus exploiting three DFT bins for the computation of the frequency deviation with respect to $f_0$. It also iteratively refines the estimate by evaluating and removing from the spectrum both the image frequency component and the others possibly present in the signal; samples are weighted with a three-cycle Hann window. For each phase, a synchrophasor and frequency estimate with an internal rate ($100\,\textrm{fps}$) are obtained. Positive sequence synchrophasor is computed and node frequency is estimated as the average of the three {\it per phase} values. ROCOF is calculated through backward discrete-time differentiation using the current and the previous internal frequency estimate. Lower $\textrm{RR}$ are obtained by decimation. 
    \paragraph{Space Vector approach (SV-F)} This method  \cite{2017TIM_Toscani_SV} is intrinsically three-phase, since it operates on the complex-valued space vector (SV) signal obtained from the three-phase samples via SV transformation. The SV signal is demodulated by adopting a reference frame that rotates at the rated frequency and processed with five different linear-phase FIR filters: the first one, extracts the baseband phasor, then its amplitude and phase angle are computed and filtered again to obtain amplitude and phase angle measurements. Frequency and ROCOF measurements are obtained applying specifically designed band-limited differentiators to the initial phase angle samples. Estimated frequency deviation is used to compensate for the scalloping loss introduced by the input filter. The performance of the method clearly depends on the filter characteristics; in the considered implementation, the design parameters reported in \cite{2017TIM_Toscani_SV} that ensure P-class compliance have been adopted: about three cycles are observed for each estimation ($N_w=601$).
    {\color{black}\paragraph{Compressive sensing weighted Taylor-Fourier multifrequency model (CS-WTFM)} The algorithm was presented in \cite{FrigPegTosWTFM}, as an enhancement of \cite{Frigo-etAl-PTC-2019}. Each component that may be present in the phase signal (i.e. fundamental, harmonic or interharmonic) is modeled with a truncated Taylor expansions, centred on the reporting instants. The frequency support, namely the components to be included in the model, are selected from a set of candidates (with $1\,\textrm{Hz}$ resolution) through a compressive sensing approach; a Chebyshev weighting window is employed to improve performance. After that, the fundamental synchrophasor and its derivatives (up to the second order) are extracted through least squares fitting of the signal model. For each phase, the synchrophasor estimate is obtained, while frequency and ROCOF values are computed relying on non-linear equations involving up to the second order derivative of the fundamental phasor, respectively. Positive sequence synchrophasor is computed, while frequency and ROCOF estimates are obtained by averaging the {\it per phase} values. A window of $N_w=3M+1$ samples around the 
    reporting instant is used, and a second order expansion for the fundamental component is adopted.}

\begin{figure}
    \centering
    \includegraphics[width=\columnwidth]{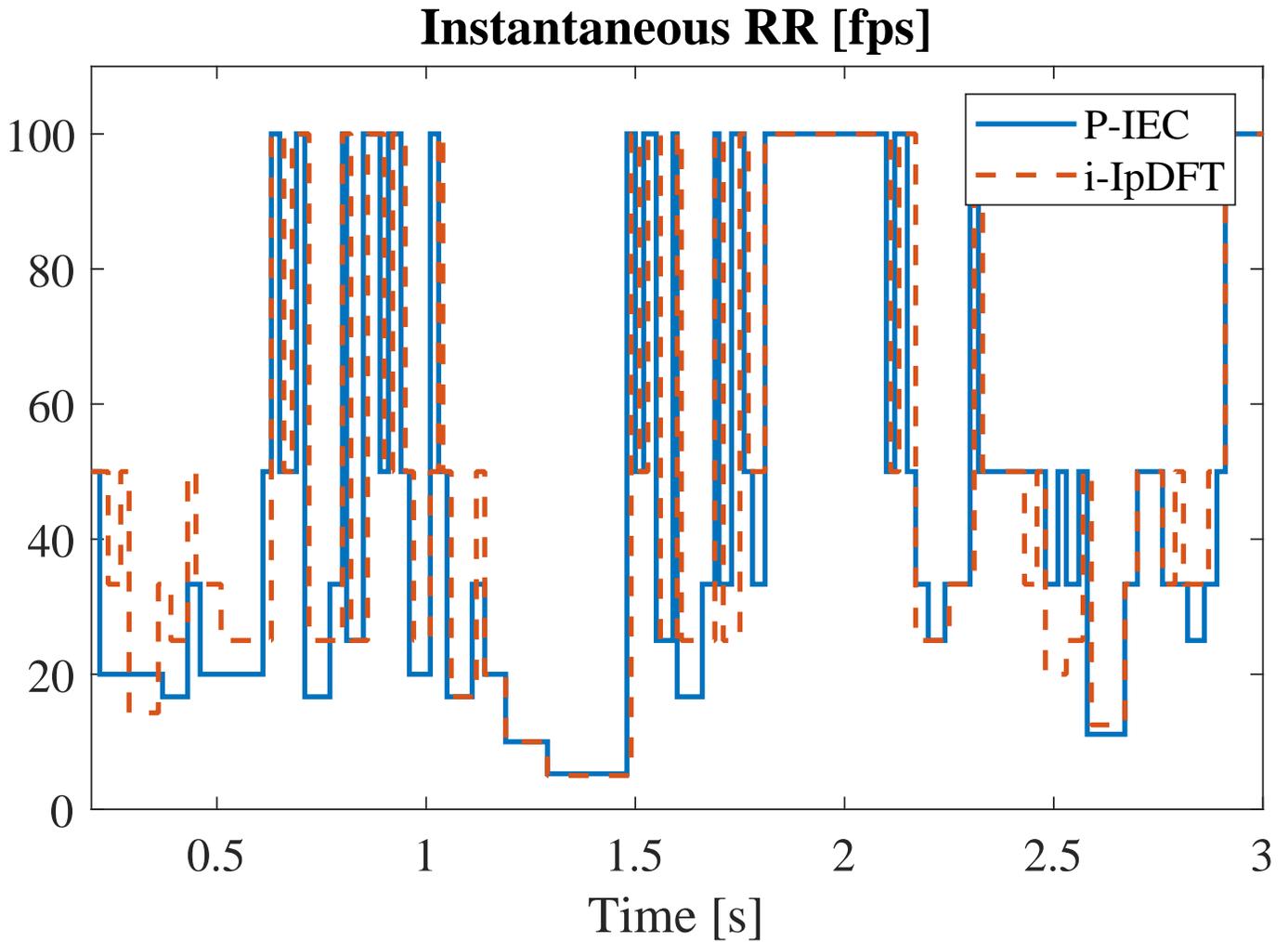}
    \caption{Australia 2016: Instantaneous RR for P-IEC and i-IpDFT algorithms.}
    \label{fig:Australia2016_instantaneousRR}
\end{figure}
\begin{figure}
    \centering
    \includegraphics[width=\columnwidth]{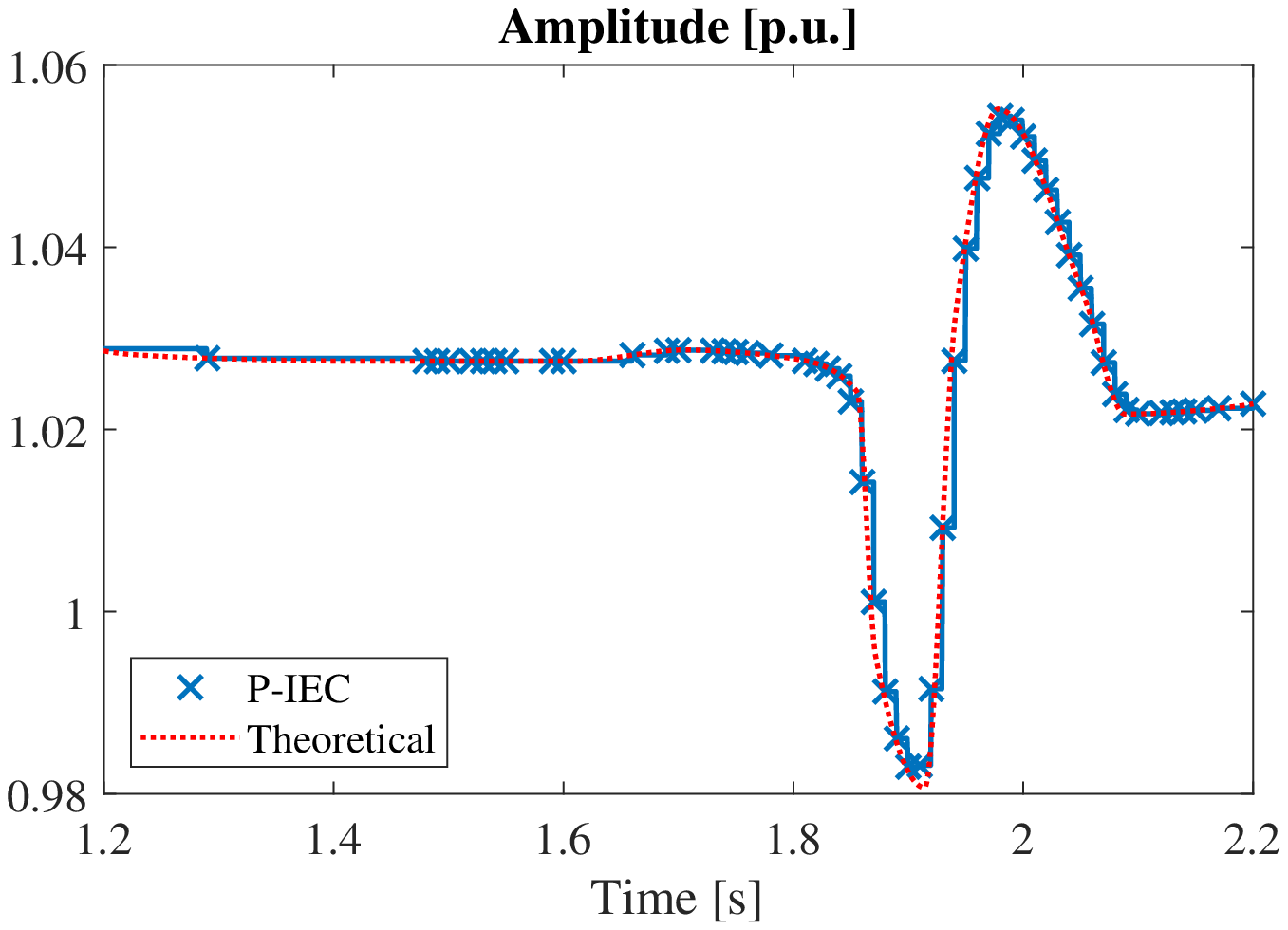}
    \caption{Australia 2016: Amplitude estimates at retained measurement instants and predicted values compared to the actual waveform, P-IEC algorithm.}
    \label{fig:Australia2016_amplitude}
\end{figure}

{\color{black}
\section{Simulation and Results}
\label{sec:results}
The test scenarios described in Section \ref{sec:case_studies} have been used to study the effectiveness of the proposed adaptive decimation, applied in conjunction with the PMU estimation algorithms summarised in Section \ref{sec:assume}. The behaviour on a relatively short waveform is firstly analysed, considering the case study Australia 2016 (Fig. \ref{fig:case_studies_short}(a)); results are reported in Table \ref{tab:TrE_FE_RFE_Australia2016}.

At $100\,\textrm{fps}$ RR, all the four algorithms achieve good phasor tracking: i-IpDFT results in $\textrm{TrE}_{\textrm{TVE}}=0.18\,\%$, while others reach just slightly lower values. When tracking errors are so close, it means that they are mostly due to the prediction rule, while the point-wise measurement accuracy of the different algorithms (which is evaluated through the conventional TVE, FE and RFE metrics) produces a negligible contribution.
Frequency tracking is also satisfactory, with i-IpDFT having the highest error ($\textrm{TrE}_{\textrm{FE}}=4.6\,\textrm{mHz}$). Similar considerations apply to $\textrm{TrE}_{\textrm{RFE}}$, with three algorithms that behave similarly, while i-IpDFT shows a tracking error that is about $50\,\%$ higher. These results are somehow expected, since i-IpDFT is inherently based on a stationary signal model, which shows its limitations as highly dynamic conditions occur. 

When applying the proposed adaptive RR approach, tracking errors slightly increase, but the variation is extremely small and not noticeable with 2 digits of precision. This is remarkable if we consider that the average RR is roughly halved, as it can be noticed from the compression ratio values, defined as the ratio between the number of measurements corresponding to $\textrm{RR}=100\,\textrm{fps}$ and those selected by the adaptive RR technique. In comparison, fixed $50\,\textrm{fps}$ reporting ratio, corresponding to a similar amount of transferred data, produces $70\,\%$ higher $\textrm{TrE}_{\textrm{TVE}}$ values, $\textrm{TrE}_{\textrm{FE}}$ may be more than double, while the increase of $\textrm{TrE}_{\textrm{RFE}}$ is between $20\,\%$ and $36\,\%$ according to the specific algorithm.

However, it is worth highlighting that i-IpDFT achieves a noticeably lower compression ratio with respect to the others. In order to better understand the behaviour, Fig. \ref{fig:Australia2016_instantaneousRR} compares the instantaneous RRs (the inverse of the time interval between consecutive selected measurements) achieved by the proposed technique as either P-IEC or i-IpDFT algorithm is adopted\footnote{From here on, the name of the algorithm in the legend of the figures without further specification indicates the application of the proposed adaptive RR algorithm to the corresponding measurements.}. The remaining two methods behave similarly to the former, therefore they are not shown for the sake of clarity.
At the beginning, the instantaneous RR is rather low, but after $0.7\,\textrm{s}$ it occasionally hits the maximum value of $100\,\textrm{fps}$ because of the frequency oscillations appearing between $0.6\,\textrm{s}$ and $1.2\,\textrm{s}$ before dropping down to about $5\,\textrm{fps}$. As a response to the abrupt variation occurring at about $1.5\,\textrm{s}$, RR increases noticeably, reaching $100\,\textrm{fps}$ for about half a second. It settles down below $50\,\textrm{fps}$, before rising again to follow the next frequency transient. From the plot, it can be noticed that the RR of the i-IpDFT is sometimes higher than that of P-IEC (and thus of the others), especially at the beginning. The reason for this is mainly related to the poor performance of the corresponding ROCOF estimate, which shows erratic oscillations, not present in the actual waveform. As a result, the threshold value $\Delta_{\textrm{RFE}}$ is exceeded more frequently, thus increasing the number of selected measurements.

Finally, Fig. \ref{fig:Australia2016_amplitude} compares the reconstructed phasor amplitude through adaptive RR with the reference one (obtained with sampling interval $T_s$) in the neighbourhood of the first large transient. The P-IEC estimation algorithm is considered here, since the others show a similar behaviour. In particular, blue crosses highlight the measurements selected by the proposed algorithm and the blue line represents the predicted values between measurements. Reported measurements are initially very sparse, but they become much denser (spacing may be as small as $10\,\textrm{ms}$) to accurately track the transient event. The zeroth order hold behaviour of the reconstructed amplitude, according to \eqref{eq:phasor_prediction}, is also evident.
}

\begin{table}
\renewcommand{\arraystretch}{1.2}
\setlength{\tabcolsep}{4pt}
    \centering
{\color{black}
    \caption{Tracking Results with Fixed and Adaptive Reporting Rate: Case Study Australia 2016}
    \begin{tabular}{cc|cccc}
    \toprule
       \multirow{2}{*}{\bf{Index}} & \bf{RR} & \multicolumn{4}{c}{\bf{Algorithm}}  \\
       & \bf{[fps]} & P-IEC &	i-IpDFT	& SV-F &	CS-WTFM\\
       \midrule
        \multirow{3}{*}{$\textrm{TrE}_\textrm{TVE}\,[\%]$} & 100 &	0.16 & 0.18 & 0.17 & 0.16\\
        & 50 & 0.30 & 0.30 & 0.30 & 0.30\\
        & adaptive & 0.17 & 0.18 & 0.17 & 0.16\\
        \midrule
        \multirow{3}{*}{$\textrm{TrE}_\textrm{FE}\,[\textrm{mHz}]$} & 100	& 2.7 & 4.6 & 3.3 & 4.3\\
        & 50	& 5.4	& 12 & 5.0 & 4.8\\
        & adaptive & 2.7 & 4.6 & 3.3 & 4.3\\
        \midrule
        \multirow{3}{*}{$\textrm{TrE}_\textrm{RFE}\,[\textrm{Hz/s}]$} & 100 & 0.80 & 1.3 & 0.81 & 0.82\\
        & 50 & 1.1 & 1.6 & 1.1 & 1.1\\
        & adaptive & 0.80 & 1.3 & 0.81 & 0.82\\
    \midrule
    Compression Ratio & - & 2.06 & 1.95 & 2.05 & 2.07\\
    \bottomrule
    \end{tabular}
    \label{tab:TrE_FE_RFE_Australia2016}
    }
\end{table}

\begin{figure}
    \centering
    \includegraphics[width=\columnwidth]{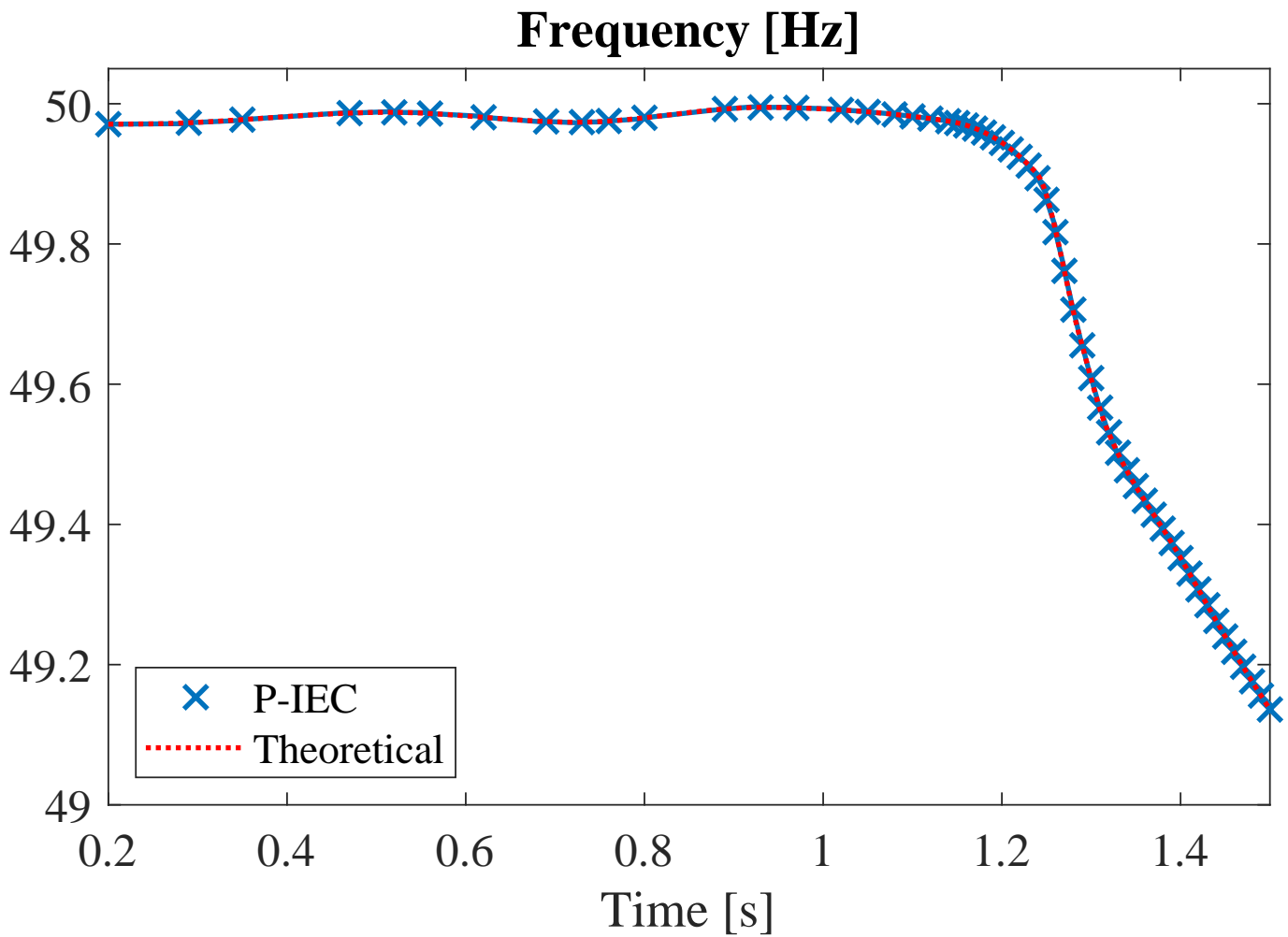}
    \caption{Arizona 2011: Frequency estimates at retained measurement instants and predicted values compared to the actual waveform, P-IEC algorithm.}
    \label{fig:Arizona2011_frequency}
\end{figure}

{\color{black}
The second, short-duration case study is represented by Arizona 2011 (Fig. \ref{fig:case_studies_short}(b)), and the most significant results are reported in Table \ref{tab:TrE_FE_RFE_Arizona2011}.
Considering fixed $100\,\textrm{fps}$ RR, synchrophasor tracking accuracy is very similar for all the considered algorithms (with i-IpDFT achieving a marginally larger $\textrm{TrE}_{\textrm{TVE}}$), and values are slightly lower than those obtained with the Australia 2016 case study. The capability to follow the time evolution of frequency is also satisfactory:  i-IpDFT reaches the largest $\textrm{TrE}_{\textrm{FE}}$, but it is still below $1\,\textrm{mHz}$. As far as the ROCOF, P-IEC, SV-F and CS-WTFM algorithms result in very close tracking errors around $0.1\,\textrm{Hz/s}$, with i-IpDFT once again being the least performing, with $\textrm{TrE}_{\textrm{RFE}}=0.23\,\textrm{Hz/s}$. Reducing RR to $50\,\textrm{fps}$ has a major effect on accuracy: the growth of the tracking error ranges from about $50\,\%$ to more than $100\,\%$. Synchrophasor tracking quality is similar for all the estimation techniques, but i-IpDFT reaches considerably larger $\textrm{TrE}_{\textrm{FE}}$ and $\textrm{TrE}_{\textrm{RFE}}$ values with respect to the others.

When adopting the proposed adaptive decimation method, we can immediately notice a minor increase of the tracking errors (below $10\,\%$) with respect to $\textrm{RR}=100\,\textrm{fps}$  for synchrophasor, frequency and ROCOF. This means that, also in this case, the set of selected measurements is able to capture virtually the full informative content. Data throughput is reduced by about $38\,\%$ for all the PMU algorithms (corresponding to an average RR of about $60\,\textrm{fps}$), thus highlighting the high efficiency of the approach. Adopting i-IpDFT results in a slightly lower compression ratio than the others, mostly because it exceeds the threshold $\Delta_{\textrm{RFE}}$ more often. As a final consideration, the adaptive decimation provides about 20\% more data with respect to the $50\,\textrm{fps}$ fixed RR, but the obtained tracking performance is far better.

Figure \ref{fig:Arizona2011_frequency} visually compares the actual frequency waveform with the prediction obtained from the selected measurements. P-IEC estimation algorithm is shown here, but the others behave very similarly. The linear trend between reported frequency measurements resulting from the prediction rule \eqref{eq:freq_prediction} is clearly visible. 
}

\begin{table}
\renewcommand{\arraystretch}{1.2}
\setlength{\tabcolsep}{4pt}
    \centering
{\color{black}
    \caption{Tracking Results with Fixed and Adaptive Reporting Rate: Case Study Arizona 2011}
    \begin{tabular}{cc|cccc}
    \toprule
       \multirow{2}{*}{\bf{Index}} & \bf{RR} & \multicolumn{4}{c}{\bf{Algorithm}}  \\
       & \bf{[fps]} & P-IEC &	i-IpDFT	& SV-F &	CS-WTFM\\
       \midrule
        \multirow{3}{*}{$\textrm{TrE}_\textrm{TVE}\,[\%]$} & 100 &	0.12 & 0.13 & 0.12 & 0.12\\
        & 50 & 0.24 & 0.24 & 0.24 & 0.24\\
        & adaptive & 0.13 & 0.14 & 0.13 & 0.12\\
        \midrule
        \multirow{3}{*}{$\textrm{TrE}_\textrm{FE}\,[\textrm{mHz}]$} & 100	& 0.56 & 0.97 & 0.56 & 0.78\\
        & 50	& 1.2	& 2.5 & 1.0 & 0.98\\
        & adaptive & 0.63 & 1.0 & 0.63 & 0.82\\
        \midrule
        \multirow{3}{*}{$\textrm{TrE}_\textrm{RFE}\,[\textrm{Hz/s}]$} & 100 & 0.10 & 0.23 & 0.10 & 0.11\\
        & 50 & 0.18 & 0.30 & 0.18 & 0.18\\
        & adaptive & 0.11 & 0.23 & 0.11 & 0.11\\
    \midrule
    Compression Ratio & - & 1.63 & 1.61 & 1.63 & 1.62\\
    \bottomrule
    \end{tabular}
    \label{tab:TrE_FE_RFE_Arizona2011}
    }
\end{table}

{\color{black}
Turkey 2015 (Fig. \ref{fig:case_studies_short}(c)) represents the last short-duration case study to be analysed. The considered segment contains a frequency ramp as well as a very jagged voltage amplitude, exhibiting large variations. When adaptive RR is employed, at least one of the threshold values is exceeded in each of the candidate reporting instant. Therefore, all the measurements are selected, and thus, in this case, the adaptive RR technique corresponds to the $100\,\textrm{fps}$ RR of the incoming data, as from Table \ref{tab:TrE_FE_RFE_Turkey2015}. Tracking errors are significantly larger than before, for example with $\textrm{TrE}_{\textrm{TVE}}$ exceeding $1\,\%$. i-IpDFT reaches the highest $\textrm{TrE}_{\textrm{FE}}$, just below $15\,\textrm{mHz}$, with the P-IEC algorithm performing just marginally better (almost $14\,\textrm{mHz}$). On the contrary, SV-F and CS-WTFM result in much lower frequency tracking errors, which drop down to about $4.5\,\textrm{mHz}$ and $3.0\,\textrm{mHz}$, respectively. Even larger differences occur as ROCOF tracking performance is analysed. Surprisingly, the P-IEC algorithm is the least accurate, with $\textrm{TrE}_{\textrm{RFE}}$ exceeding $1.4\,\textrm{Hz/s}$, while i-IpDFT enables an error that is $40\,\%$ lower. However, SV-F and CS-WTFM methods provide a further significant improvement, with ROCOF tracking errors of $0.47\,\textrm{Hz/s}$ and $0.15\,\textrm{Hz/s}$, respectively. These results show that, when needed, i.e., when dynamics to track require a finer-grained measurement sampling, the proposed algorithm behaves exactly as expected, without discarding any relevant information.  
}

\begin{table}
\renewcommand{\arraystretch}{1.2}
\setlength{\tabcolsep}{4pt}
    \centering
{\color{black}
    \caption{Tracking Results with Fixed and Adaptive Reporting Rate: Case Study Turkey 2015}
    \begin{tabular}{cc|cccc}
    \toprule
       \multirow{2}{*}{\bf{Index}} & \bf{RR} & \multicolumn{4}{c}{\bf{Algorithm}}  \\
       & \bf{[fps]} & P-IEC &	i-IpDFT	& SV-F &	CS-WTFM\\
       \midrule
        \multirow{2}{*}{$\textrm{TrE}_\textrm{TVE}\,[\%]$} & 100 &	1.1 & 1.2 & 1.1 & 1.1\\
        & adaptive & 1.1 & 1.2 & 1.1 & 1.1\\
        \midrule
        \multirow{2}{*}{$\textrm{TrE}_\textrm{FE}\,[\textrm{mHz}]$} & 100	& 13.7 & 14.9 & 4.5 & 3.0\\
        & adaptive & 13.7 & 14.9 & 4.5 & 3.0\\
        \midrule
        \multirow{2}{*}{$\textrm{TrE}_\textrm{RFE}\,[\textrm{Hz/s}]$} & 100 & 1.4 & 0.90 & 0.47 & 0.15\\
        & adaptive & 1.4 & 0.90 & 0.47 & 0.15\\
    \midrule
    Compression Ratio & - & 1.00 & 1.00 & 1.00 & 1.00\\
    \bottomrule
    \end{tabular}
    \label{tab:TrE_FE_RFE_Turkey2015}
    }
\end{table}

{\color{black}

The second set of tests is focused on longer waveforms, i.e., Florida 2019 and Croatia 2021 events presented in Section \ref{sec:case_studies}, aiming at assessing the performance of the proposed method also on a wider time interval where either fast or long-term events can occur. 

Figure \ref{fig:Florida2019_frequency} shows frequency measurements in a 15-s portion of Florida 2021. 
Once again, the tracking appears significantly accurate with respect to sample-by-sample reference values. As expected, the algorithm output focuses more on faster dynamics, whereas the measurement instants start to thin out as the frequency transition becomes smoother. Frequency follows an oscillatory behaviour all over the waveform (see Fig. \ref{fig:case_studies_long}(a)) and thus the compression pattern appears somehow periodical too. This is confirmed by Fig. \ref{fig:Florida2019_instantaneous_RR}, which reports the instantaneous RR provided by the proposed algorithm during the whole event. It is interesting to notice that the instantaneous RR is much lower than the original one (100~fps), leading to an average RR of $7.1\,\textrm{fps}$ and thus to a remarkable compression ratio.

\begin{figure}
    \centering
    \includegraphics[width=\columnwidth]{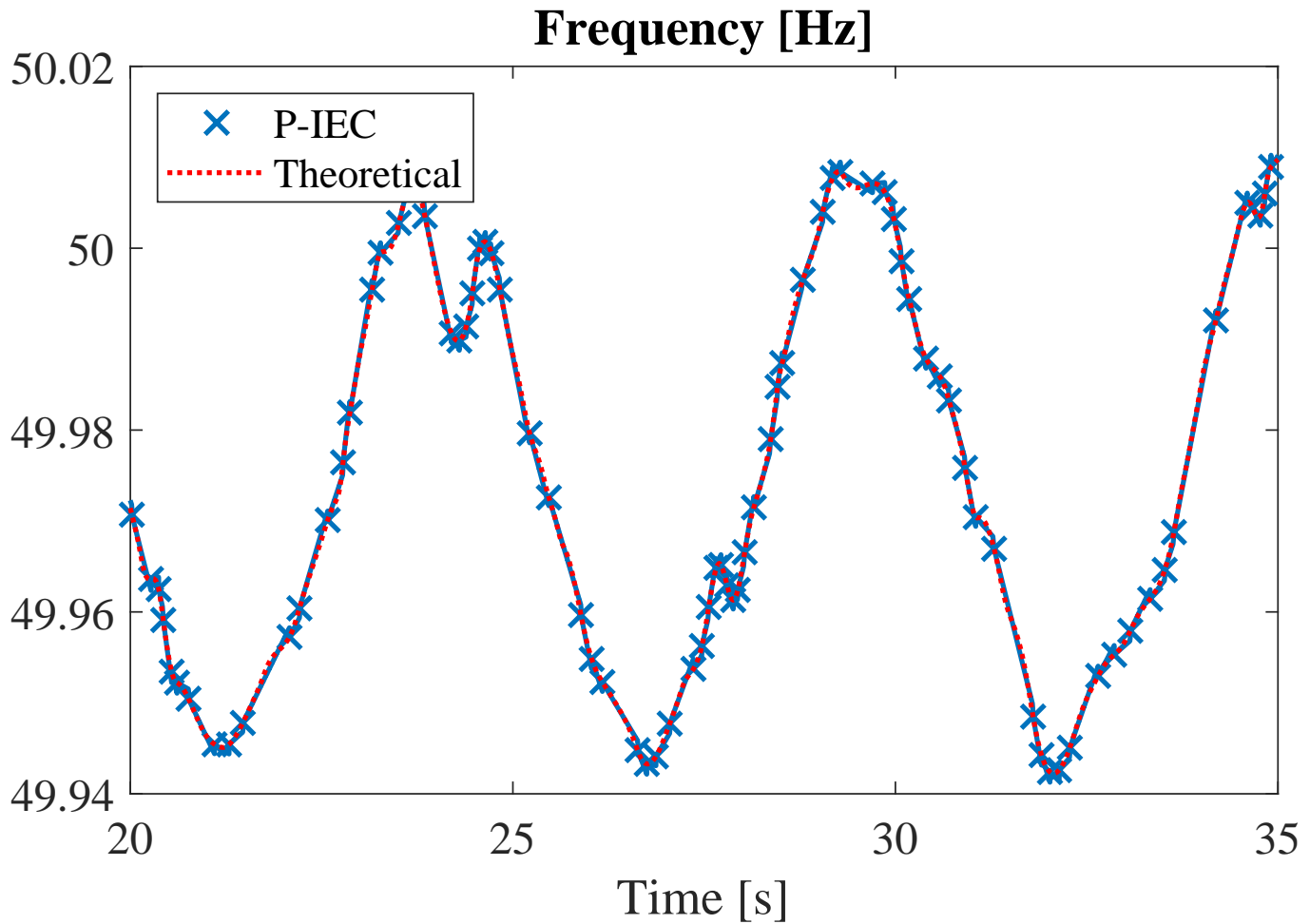}
    \caption{Florida 2019: Frequency estimates at retained measurement instants and predicted values compared to the actual waveform, P-IEC algorithm.}
    \label{fig:Florida2019_frequency}
\end{figure}

\begin{figure}
    \centering
    \includegraphics[width=\columnwidth]{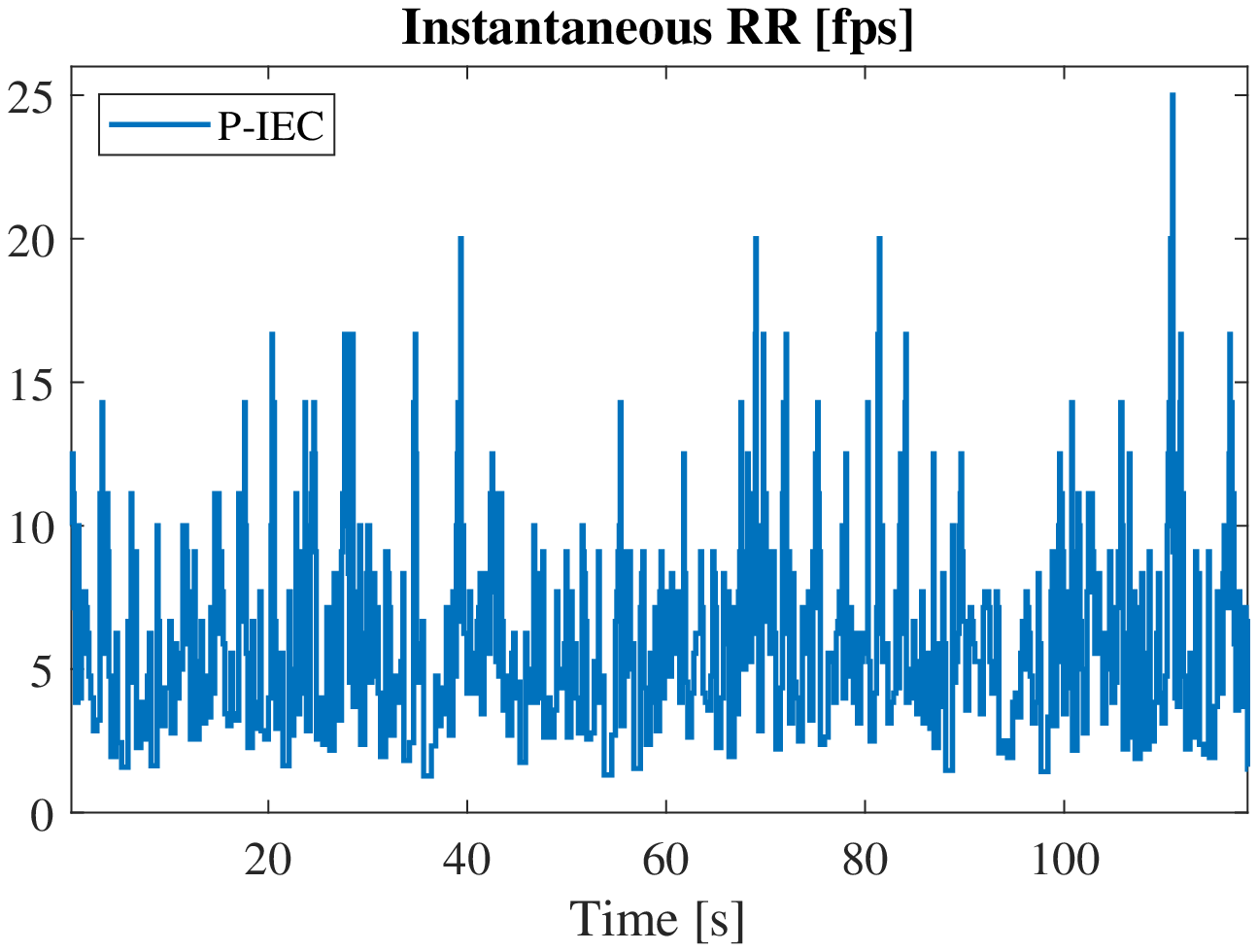}
    \caption{Florida 2019: Instantaneous RR, P-IEC algorithm.}
    \label{fig:Florida2019_instantaneous_RR}
\end{figure}

Table \ref{tab:TrE_FE_RFE_Florida2019} summarises the tracking results for Florida 2019, with all the considered indices. Compared to the original $100\,\textrm{fps}$ measurement series, the adaptive RR leads to a compression ratio above 17, corresponding to less than 700 selected estimates out of about 12000. The tracking indices for TVE, FE and RFE confirm the capability of the method to maintain low prediction errors, while reducing the amount of data. It is important to underline that the aim of the proposal is to bound the errors within reasonable and desired limits for the whole time rather than focusing on maximum accuracy. Nevertheless, by suitably tuning the thresholds, it is still possible to increase the PMU accuracy at the expenses of higher data rates.

Table \ref{tab:TrE_FE_RFE_Florida2019} also reports the results for $5\,\textrm{fps}$, since such RR roughly corresponds to the average measurement rate of the adaptive method. The proposed approach results in much lower errors with a similar amount of data.    
$\textrm{TrE}_\textrm{FE}$ is indeed reduced to almost one third and is kept lower than $1\,\textrm{mHz}$ for all the algorithms. It is important to stress that, with such a periodical time evolution of the signal parameters, we are in a favourable condition for constant RRs. In practice, long steady-state and close to nominal conditions are much more common than dynamics and thus the advantage of the adaptive RR is even more pronounced.  

It is interesting to notice that the different algorithms behave in a similar way, i.e., they show similar tracking capabilities under adaptive RR, whereas constantly high RRs tend to emphasise the differences. With the original data rate, in particular, errors are low but fairly different between the methods. For instance, i-IpDFT, which suffers from larger FEs during dynamic conditions, results in higher tracking error indices for all the measured quantities.

}

\begin{table}
\renewcommand{\arraystretch}{1.2}
\setlength{\tabcolsep}{4pt}
    \centering
{\color{black}
    \caption{Tracking Results with Fixed and Adaptive Reporting Rate: Case Study Florida  2019}
    \begin{tabular}{cc|cccc}
    \toprule
       \multirow{2}{*}{\bf{Index}} & \bf{RR} & \multicolumn{4}{c}{\bf{Algorithm}}  \\
       & \bf{[fps]} & P-IEC &	i-IpDFT	& SV-F &	CS-WTFM\\
       \midrule
        \multirow{3}{*}{$\textrm{TrE}_\textrm{TVE}\,[\%]$} & 100 &	6.5 E-4	& 11 E-4 & 9.0 E-4	& 2.0 E-4\\
        & 5 &	5.2 E-2 & 6.0 E-2 & 5.2 E-2 & 5.2 E-2\\
        & adaptive & 2.6 E-2 & 2.5 E-2 & 2.5 E-2 & 2.5 E-2\\
        \midrule
        \multirow{3}{*}{$\textrm{TrE}_\textrm{FE}\,[\textrm{mHz}]$} & 100	& 0.005 & 0.011 & 0.007	& 0.010\\
        & 5	& 1.3	& 1.5 & 1.3 & 1.3\\
        & adaptive & 0.46 & 0.47 & 0.46 & 0.46\\
        \midrule
        \multirow{3}{*}{$\textrm{TrE}_\textrm{RFE}\,[\textrm{Hz/s}]$} & 100 & 0.001 & 0.003 & 0.001 & 0.001\\
        & 5 & 0.016	& 0.017	& 0.016 & 0.016\\
        & adaptive & 0.010 & 0.010 & 0.010 & 0.010\\
    \midrule
    Compression Ratio & - & 18.8 & 17.7 & 19.0 & 19.1\\
    \bottomrule
    \end{tabular}
    \label{tab:TrE_FE_RFE_Florida2019}
    }
\end{table}

\begin{figure}
    \centering
    \includegraphics[width=\columnwidth]{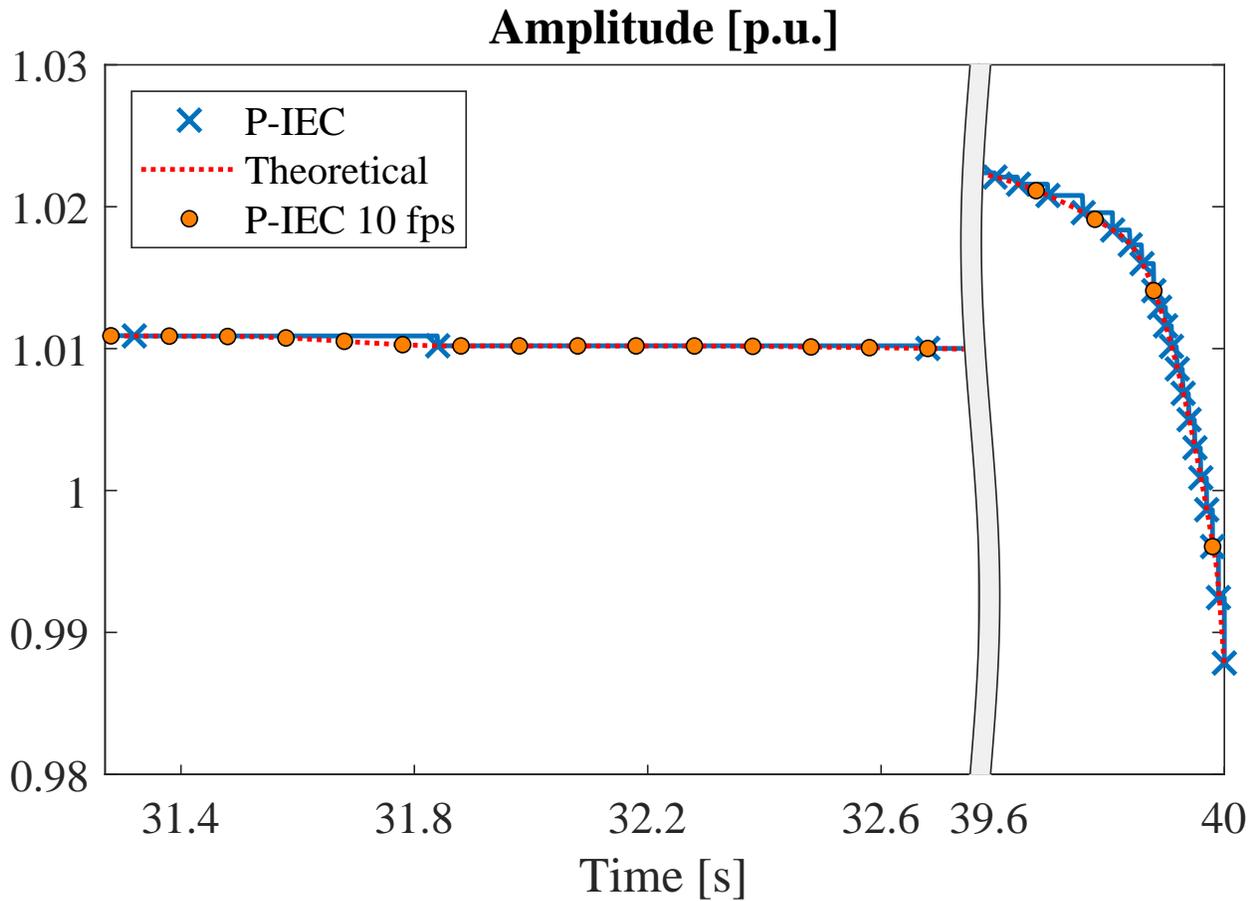}
    \caption{Croatia 2021: 
    Phasor amplitude estimation at retained measurement instants with adaptive and fixed-rate decimation.}
    \label{fig:Croatia2021_amplitude}
\end{figure}

{\color{black}
The Croatia 2021 waveform includes a much more irregular behaviour with respect to Florida 2019, particularly for voltage magnitude (see Fig. \ref{fig:case_studies_long}(b)), with fast oscillations around $t=40\,\textrm{s}$. The proposed method has been applied with the same parameters as before, and the tracking results are reported in Table \ref{tab:TrE_FE_RFE_Croatia2021}, with three different RRs: $100\,\textrm{fps}$, $10\,\textrm{fps}$, and adaptive. $10\,\textrm{fps}$ is the closest divider of $\textrm{RR}_{\textrm{in}}$ to the average RR corresponding to the compression ratio of the adaptive approach, which is higher than 10. The results show that in such conditions, $\textrm{TrE}_\textrm{TVE}$ degradation is much stronger (almost 10 times) when a fixed decimation rate is applied with respect to the adaptive case. This behaviour can be understood by looking at Fig. \ref{fig:Croatia2021_amplitude}, where the adaptive sampling method is compared with the $\textrm{RR}=10\,\textrm{fps}$ case in terms of magnitude measurements. Two different time intervals have been magnified in the figure. The left inset refers to an almost steady-state condition around $t=32\,\textrm{s}$. It is clear that the adaptive method (blue crosses) gives only few relevant measurements that are added when the zero-holding of the previous magnitude estimate is not enough to properly track the variations. On the other side, the inset on the right in Fig. \ref{fig:Croatia2021_amplitude} shows that the adaptive method chooses frequent measurements to follow fast dynamics, whereas measurements performed every $100\,\textrm{ms}$ (orange points) easily miss the fast transition. In the figure, for the sake of clarity, only predictions computed through the adaptive rate are reported (which are very close to the theoretical dashed line), but it is easy to understand that zero-holding in the $\textrm{RR}=10\,\textrm{fps}$ case (not shown for a better clarity) leads to large errors just before $t=40\,\textrm{s}$.
Visual inspection of Fig. \ref{fig:Croatia2021_amplitude} also helps  
to recall that tracking indices have an averaging effect in case of fast changes having short duration, and thus the overall error decrease summarised in Table~\ref{tab:TrE_FE_RFE_Croatia2021} is even more significant when maximum errors or average errors in small intervals are considered.
}

\begin{table}
\renewcommand{\arraystretch}{1.2}
\setlength{\tabcolsep}{4pt}
    \centering
{\color{black}
    \caption{Tracking Results with Fixed and Adaptive Reporting Rate: Case Study Croatia 2021}
    \begin{tabular}{cc|cccc}
    \toprule
       \multirow{2}{*}{\bf{Index}} & \bf{RR} & \multicolumn{4}{c}{\bf{Algorithm}}  \\
       & \bf{[fps]} & P-IEC &	i-IpDFT	& SV-F &	CS-WTFM\\
       \midrule
        \multirow{3}{*}{$\textrm{TrE}_\textrm{TVE}\,[\%]$} & 100 & 0.020 & 0.020 & 0.020 & 0.020\\
                                                           & 10  & 0.19 & 0.19 & 0.19 & 0.19\\
                                                      & adaptive & 0.05 & 0.05 & 0.05 & 0.05\\
        \midrule
        \multirow{3}{*}{$\textrm{TrE}_\textrm{FE}\,[\textrm{mHz}]$} & 100 & 0.038 &	0.044 &	0.015 &	0.049\\
                                                                    & 10  & 0.41 &	0.48 & 0.36 & 0.40\\
                                                               & adaptive & 0.35 &	0.37 &	0.35 &	0.35\\
        \midrule
        \multirow{3}{*}{$\textrm{TrE}_\textrm{RFE}\,[\textrm{Hz/s}]$} & 100 & 0.004	& 0.004 & 0.002 & 0.002\\
                                                                      & 10  & 0.010 & 0.011 & 0.010 & 0.010\\
                                                                 & adaptive & 0.008 & 0.008 & 0.007 & 0.007\\
    \midrule
    Compression Ratio & - & 10.7 & 10.5 & 10.7 & 10.7\\
    \bottomrule
    \end{tabular}
    \label{tab:TrE_FE_RFE_Croatia2021}
    }
\end{table}

\section{Conclusions}
\label{sec:concl}
{\color{black}The paper has addressed the problem of reducing measurement data rate from PMUs while keeping the relevant information, so that it is possible to reconstruct the time evolution of the monitored quantities according to a predetermined accuracy target. In particular, the proposed method continuously adapts the measurement reporting rate of synchrophasor, frequency and ROCOF to actual signal conditions. Thanks to this capability, it can effectively cope with both events requiring more detailed descriptions and steady-state conditions that can be summarised with few measurements.
Waveforms from different real-world scenarios have been used to validate the algorithm and prove its efficiency and performance. Indeed the tracking accuracies based on selected measurements are comparable with those at maximum reporting rate and, on the other hand, they are remarkably better than those achievable with constant reporting rates providing the same data throughput. 

Critical events can be followed in an efficient way, and different PMU algorithms (e.g., from different PMU models) can be adapted to achieve similar tracking performance targets, thus highlighting the importance of focusing on application objectives instead than individual instrument performance.
The proposed method can thus represent a valuable tool for transmission system operators to configure their wide area monitoring systems and enhance its operation, according to their specific needs.}

\bibliographystyle{IEEEtran}
\bibliography{IEEEabrv.bib,biblio.bib}

\end{document}